\RequirePackage{lineno}
\documentclass[prl,twocolumn,showpacs,amsmath,amssymb]{revtex4-1}
\usepackage{graphicx}
\usepackage{dcolumn}
\usepackage{bm}
\usepackage{rotating}
\usepackage{epstopdf}
\usepackage{color}
\usepackage{verbatim} 
\usepackage{multirow}
\usepackage[abs]{overpic}
\usepackage{amsmath}
\usepackage{mathrsfs}
\usepackage{amssymb}
\usepackage{subfigure}
\usepackage{xspace}
\usepackage{float}
\usepackage[colorlinks,
            linkcolor=blue,
            anchorcolor=blue,
            citecolor=blue]{hyperref}

\newcommand{\PreserveBackslash}[1]{\let\temp=\\#1\let\\=\temp}
\newcolumntype{C}[1]{>{\PreserveBackslash\centering}p{#1}}
\newcolumntype{R}[1]{>{\PreserveBackslash\raggedleft}p{#1}}
\newcolumntype{L}[1]{>{\PreserveBackslash\raggedright}p{#1}}

\newcommand{\EE}{e^+e^-}

\newcommand{\jpsi}{J/\psi}
\newcommand{\piz}{\pi^{0}}

\newcommand{\too}{\rightarrow}

\begin{document}
\graphicspath{{figure/}}
\DeclareGraphicsExtensions{.eps,.png,.ps}
\title{Observation of Charmonium $h_{c}$ Radiative Decays to Multiple Light Hadrons and the Tensor state $f_2(1270)$}

\author{
\begin{small}
\begin{center}
M.~Ablikim$^{1}$, M.~N.~Achasov$^{4,c}$, P.~Adlarson$^{76}$, X.~C.~Ai$^{81}$, R.~Aliberti$^{35}$, A.~Amoroso$^{75A,75C}$, Q.~An$^{72,58,a}$, Y.~Bai$^{57}$, O.~Bakina$^{36}$, Y.~Ban$^{46,h}$
% ...
\\
\vspace{0.2cm}
(BESIII Collaboration)\\
\vspace{0.2cm} {\it
$^{1}$ Institute of High Energy Physics, Beijing 100049, People's Republic of China\\
% ...
$^{4}$ Budker Institute of Nuclear Physics SB RAS (BINP), Novosibirsk 630090, Russia\\
% ...
$^{35}$ Johannes Gutenberg University of Mainz, Johann-Joachim-Becher-Weg 45, D-55099 Mainz, Germany\\
$^{36}$ Joint Institute for Nuclear Research, 141980 Dubna, Moscow region, Russia\\
% ...
$^{46}$ Peking University, Beijing 100871, People's Republic of China\\
% ...
$^{57}$ Southeast University, Nanjing 211100, People's Republic of China\\
$^{58}$ State Key Laboratory of Particle Detection and Electronics, Beijing 100049, Hefei 230026, People's Republic of China\\
% ...
$^{72}$ University of Science and Technology of China, Hefei 230026, People's Republic of China\\
% ...
$^{75}$ University of Turin and INFN, (A)University of Turin, I-10125, Turin, Italy; (B)University of Eastern Piedmont, I-15121, Alessandria, Italy; (C)INFN, I-10125, Turin, Italy\\
$^{76}$ Uppsala University, Box 516, SE-75120 Uppsala, Sweden\\
% ...
$^{81}$ Zhengzhou University, Zhengzhou 450001, People's Republic of China\\
\vspace{0.2cm}
$^{a}$ Deceased\\
% ...
$^{c}$ Also at the Novosibirsk State University, Novosibirsk, 630090, Russia\\
% ...
$^{h}$ Also at State Key Laboratory of Nuclear Physics and Technology, Peking University, Beijing 100871, People's Republic of China\\
% ...
}
\end{center}
\vspace{0.4cm}                                                                         
\end{small}
}

\begin{abstract} Using $\psi(3686)\rightarrow\pi^{0} h_{c}$ decays from a data
sample of $(27.12\pm0.14)\times10^{8}$ $\psi(3686)$ events collected
by the BESIII detector at the BEPCII collider, $h_c$ radiative decays
to multiple light hadrons below 2.8 GeV/$c^2$,
$\gamma\pi^{+}\pi^{-},~\gamma\pi^{+}\pi^{-}\eta,~\gamma2(\pi^{+}\pi^{-})$,
and $\gamma p\bar{p}$ are observed for the first time, each with a
significance greater than $5\sigma$. The corresponding branching
fractions are measured. Furthermore, intermediate states are investigated, leading to the first observation of the
decay process of $h_c\rightarrow\gamma
f_{2}(1270)\rightarrow\gamma\pi^{+}\pi^{-}$ with a significance of
$5.1\,\sigma$. This observation represents the first instance of $h_c$
radiative decay to a tensor state. \end{abstract}


\maketitle

The study of charmonium radiative decay provides a unique opportunity
to deepen our understanding of quantum chromodynamics (QCD) by testing
theoretical predictions with experimental observations. Of particular
interest, enhanced glueball production is expected in OZI-suppressed
processes, such as $J/\psi$ radiative decay, where the $c\bar{c}$ must
annihilate to gluons before hadronizing into the final state.  The
radiative decays of the P-wave triplet charmonium
($\chi_{cJ}$)~\cite{CLEO:2008sah,BESIII:2011ysp} have been
investigated to test our theoretical understanding of their decay
dynamics~\cite{Gao:2006bc}. However, the study of the radiative decay
of the P-wave singlet $h_{c}$ charmonium state to light hadrons is
limited due to its predominantly suppressed production through
$\psi(3686)$~\cite{BESIII:2022tfo}. Only three $h_c\too\gamma+1
^{1}S_{0}$ decays have been observed, namely
$h_{c}\too\gamma\eta_{c}$~\cite{pdg}, $h_{c}\too\gamma\eta^{\prime}$,
and $h_{c}\too\gamma\eta$~\cite{hcgamP}. The branching fraction
$\mathcal{B}(h_c\too\gamma + 2g)$, where $g$ represents gluon, is
estimated to be approximately 5.5\% with a QCD
model~\cite{qcdhc}. However, the measured total branching fraction of
$h_{c}$ decay to a photon and a light hadron($\eta/\eta^{\prime}$) is
only about 0.2\%~\cite{hcgamP}, indicating that a substantial portion
of $h_{c}\too\gamma+ 2g$ decay modes remain to be seen.

Studying $\jpsi$ and $\psi(3686)$ radiative
photon recoil spectra and comparing their decay rates has proven to be
quite informative~\cite{gam2pi, gam2pieta, gam4pi, gamppbar, jpsisearch1,
jpsisearch2, psipsearch}. Many channels, including
$\gamma\pi^{+}\pi^{-},~\gamma\pi^{+}\pi^{-}\eta,~\gamma2(\pi^{+}\pi^{-})$,
and $\gamma p\bar{p}$, have been studied to search for exotic
states~\cite{gam2pi, gam2pieta, gam4pi, gamppbar}, such as the
$p\bar{p}$ near-threshold invariant-mass enhancement in $J/\psi\too\gamma p
\bar{p}$~\cite{gamppbar}. Among these, the $f_{2}(1270)$ state has
generated particular interest, with various theoretical models being
proposed for its description, including the standard $q\bar{q}$ pair
model, a $\rho-\rho$ molecule~\cite{rho2model}, or a mixture between the
$q\bar{q}$ component and the glueball component~\cite{qqgmodel}. The
$h_{c}$ has the same C-parity as the $\jpsi$ but different orbital
angular momentum,
and its unique properties
provide an opportunity for improving our understanding of charmonium
radiative decays and for searching for exotic particles.

In this Letter, we present the first measurements of four $h_c$
radiative decays into multiple light hadrons, based on a dataset of
$(27.12\pm0.14)\times10^{8}$ $\psi(3686)$ events~\cite{0912data}
collected by the BESIII detector~\cite{besiii}. We measure the
branching fractions of radiative decays to I:~$\pi^{+}\pi^{-}$,
II:~$\pi^{+}\pi^{-}\eta$, III:~$2(\pi^{+}\pi^{-})$, and IV:~$p\bar{p}$
in radiative decays and investigate intermediate states to search for
potential exotic states.

The BESIII detector records symmetric $\EE$ collisions at the BEPCII
collider~\cite{bepcii}. Details of the BESIII detector can be found in
Ref.~\cite{besiii}. Simulated data samples produced with {\sc
geant4}-based~\cite{geant4} Monte Carlo (MC) software, which includes
the geometric description of the BESIII detector and the detector
response, are used to determine the detection efficiency and to
estimate the background contributions. The simulation includes the
beam energy spread and initial-state radiation (ISR) in the $e^+e^-$
annihilations modeled with the generator {\sc kkmc}~\cite{KKMC}. The
inclusive MC sample includes the production of the $\psi(3686)$
resonance, the ISR production of the $\jpsi$, and the continuum
processes incorporated in {\sc kkmc}~\cite{KKMC}. All particle decays
are modeled with {\sc evtgen}~\cite{ref:evtgen} using branching
fractions either taken from the Particle Data Group (PDG)~\cite{pdg},
when available, or otherwise estimated with {\sc
lundcharm}~\cite{ref:lundcharm}. Final state radiation from charged
final state particles is incorporated using {\sc
photos}~\cite{photos}.  The signal MC samples for the decay
$\psi(3686)\to\piz h_{c}$ are generated using a helicity amplitude
model. The $h_c$ exclusive decays are simulated using a phase
space (PHSP) model~\cite{ref:evtgen}, except for
the process of $h_{c}\too\gamma f_{2}(1270)$, which is modeled with an angular distribution of $1+\cos^{2}(\theta)$, where
$\theta$ is the polar angle of the radiative photon in the rest frame
of the $h_{c}$. Subsequently, $f_{2}(1270)\too\pi^{+}\pi^{-}$ is
generated according to the TSS model~\cite{ref:evtgen}, which
describes the decay of a tensor particle into a pair of scalar mesons.

Charged tracks and photon candidates are selected based on standard
BESIII criteria~\cite{BESIII:2022olx}. 
The number of candidate charged tracks must exactly match the number of charged particles in final states, while the number of photons must be no less than the number required by each decay mode.
The $h_c$ meson is produced by the
isospin-violating decay $\psi(3686) \rightarrow \pi^{0}h_c$, and the
$\piz$ mass acceptance window is set to be $[0.10,~0.16]\,{\rm
  GeV}/c^{2}$ in the event pre-selection. To reduce background
contributions and improve the mass resolution, a five-constraint (5C)
kinematic fit is performed with four constraints (4C) requiring that the
total four-momentum of final states is equal to the initial
$\psi(3686)$ momentum and an additional $\piz$ mass~\cite{pdg}
constraint. The one with the minimum $\chi^{2}_{\text{5C}}$ value is selected in all possible photon combinations. For the $\psi(3686)\too\piz h_{c}$,
$h_{c}\too\gamma\pi^{+}\pi^{-}\eta$ process where the
$\eta\too\gamma\gamma$ decay is utilized, at least five photons are
required. The two photons with an invariant mass closest to the $\eta$
mass~\cite{pdg} are taken as originating from the $\eta$. The $\eta$
signal region is defined as [0.513, 0.583] GeV/$c^{2}$, while the
sideband regions, [0.400, 0.470] and [0.626, 0.696] GeV/$c^{2}$, are
used to estimate the non-$\eta$ background. For the process of
$h_{c}\too\gamma p\bar{p}$, particle identification (PID) is used for
the charged particles. The flight time in the time-of-flight system
(TOF) and the d$E$/d$x$ information in the multilayer drift chamber
(MDC) are combined to calculate PID likelihoods for the $\pi$, $K$,
and $p$ hypotheses. For this channel, the likelihood of a proton
hypothesis is required to be larger than those for both the kaon and
pion hypotheses.

To suppress backgrounds with the wrong number of photons compared to
the signal mode, $\chi^{2}_{4C,n\gamma}<\chi^{2}_{4C,(n-1)\gamma}$ and
$\chi^{2}_{4C,n\gamma}<\chi^{2}_{4C,(n+1)\gamma}$ are required for Modes
I, III, and IV. For Mode II, only
$\chi^{2}_{4C,n\gamma}<\chi^{2}_{4C,(n+1)\gamma}$ is used, because the
inclusion of $\chi^{2}_{4C,n\gamma}<\chi^{2}_{4C,(n-1)\gamma}$ does
not significantly improve the purity of the signal. Here
$\chi^{2}_{4C,n\gamma}$ is the $\chi^{2}$ value of the
4C kinematic fit with the hypothesis of the same photon combination selected with the minimum $\chi^2_{5C}$, where $n$ includes photons from $\piz$ and $h_c$ decays in the process of $\psi(3686)\too\piz h_{c}$. And $\chi^{2}_{4C,(n-1)\gamma}$ and
$\chi^{2}_{4C,(n+1)\gamma}$ are the minimum $\chi^{2}$ values obtained from the 4C fits method with one less and one more photon compared to the signal process, respectively.

To reduce the peaking background of $h_{c}\too\gamma\eta_{c}$ with
$\eta_{c}\rightarrow X$ and $X$ are $\pi^+ \pi^-\eta$,
$\pi^+ \pi^- \pi^+ \pi^-$, and $p \bar{p}$, the invariant mass of $X$
is required to be less than 2.8 GeV/$c^{2}$. Additionally, for Mode
II, $M(\pi^{+}\pi^{-}\eta)>1.0$ GeV/$c^{2}$ is required to reject the
peaking background from $h_{c}\too\gamma\eta^{\prime}$.  After
applying all the above selection criteria, two types of peaking
backgrounds remain: $h_c\too\piz+X$ with a $\gamma$ missing from the
$\piz$ and $h_c\too\gamma\eta_{c}$ (with $\eta_{c}\too X$ and $M(X)<$
2.8 GeV/$c^{2}$). The contribution from the first type is estimated
using a MC simulation according to the measured $h_c \to \piz X$
branching fractions~\cite{hc3pi, hc3pieta}. For the second one, the
contribution of $h_c\too\gamma\eta_{c}$ is obtained by fitting $M(X)$
in the range [2.40, 3.07] GeV/$c^{2}$. A Breit-Wigner function
convolved with a Gaussian function is used to describe the signal
yield in the fit, with parameters fixed to the $\eta_{c}$ nominal
values~\cite{pdg}. The smooth background shape is modeled by a second-order polynomial function. The number of remaining $\eta_{c}$ events with
$M(X)<$ 2.8 GeV/$c^{2}$ is then determined to be less than 2\%. For
Mode (II), there is no peaking background from the $\eta$ sideband
region.

For non-peaking backgrounds, the following specific requirements are
imposed. Mode I: the criterion $\cos\theta_{\pi^{+}\pi^{-}}>-0.98$,
where $\theta_{\pi^{+}\pi^{-}}$ is the angle between the two pions, is
used to suppress the processes of two leptons produced by the $\EE$ collision. The invariant mass of $\pi^{+}\pi^{-}\pi^{0}$ is required to be outside the $\omega$ and $\phi$ mass regions by 3$\sigma$, where $\sigma$ is the corresponding invariant mass resolution, with values of 9.8 and 9.1 MeV/$c^2$ for $\omega$ and $\phi$, respectively.
Also $M(\pi^{0}\gamma)$ is required to be outside the $\omega$ mass region by 3$\sigma$($\sigma=12.1$ MeV/$c^2$) to reject the background from
$\omega\too\pi^{0}\gamma$. To remove photon combinatorial background,
$M(\gamma_{H}\gamma)$ is required to be outside the $\piz$ mass window $[0.10,~0.16]\,{\rm GeV}/c^{2}$, where $\gamma_{H}$ is the higher-energy
photon from $\piz$ decay and $\gamma$ is the photon from $h_c$
radiative decay. The background from $\psi(3686)\too\piz\piz J/\psi$ is suppressed by $M(\pi^+ \pi^-)<2.8 \,{\rm GeV}/c^{2}$. Mode II: photon combinatorial background is removed by requiring that $M(\gamma\gamma)_{\rm mix}$ does not fall within the
$\piz$ mass window, where $M(\gamma\gamma)_{\rm mix}$ denotes the
invariant mass of all possible two photon pairs, which are not from
the same $\pi^{0}$ or $\eta$ candidate. Events with $\pi^{+}\pi^{-}$
recoil mass greater than 3.085 GeV/$c^{2}$ are removed to veto the
background from $\psi(3686)\too \pi^{+}\pi^{-}J/\psi$. Mode III: the
$\pi^{+}\pi^{-}\pi^{0}$ invariant mass must be outside the $\eta$ ($[0.537,~0.558]\,{\rm GeV}/c^{2}$) and
$\omega$ ($[0.762,~0.804]\,{\rm GeV}/c^{2}$) mass regions, and the $\pi^{+}\pi^{-}$ recoil mass should not
fall within the $J/\psi$ mass region $[3.091,~3.103]\,{\rm GeV}/c^{2}$.

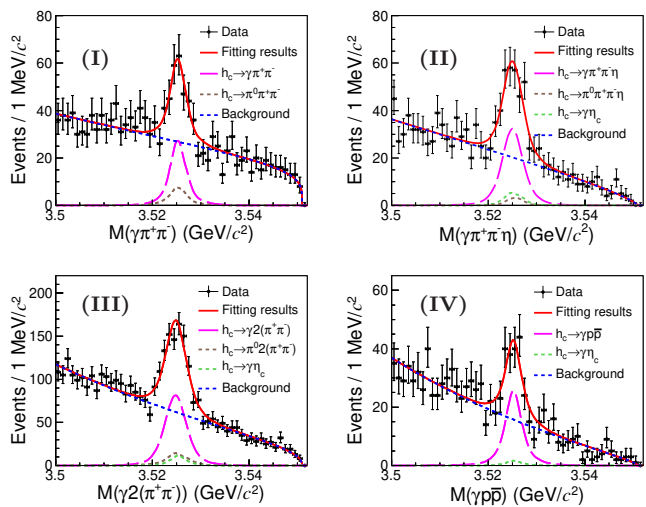
\begin{figure}[htbp]
\begin{center}
\begin{overpic}[width=0.45\textwidth, trim=35 30 20 20]{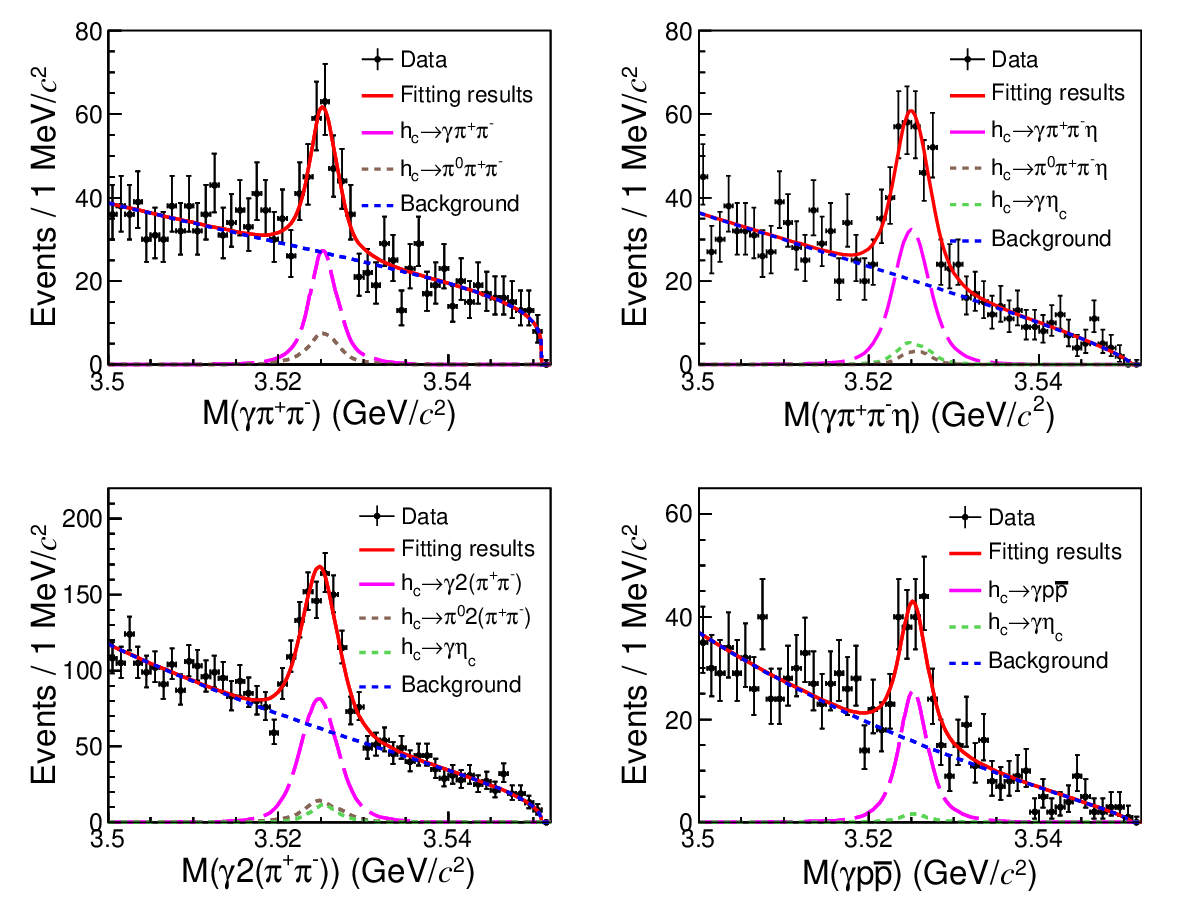}
\put(17,158){\footnotesize{\textbf{(I)}}}
\put(145,158){\footnotesize{\textbf{(II)}}}
\put(17,65){\footnotesize{\textbf{(III)}}}
\put(145,65){\footnotesize{\textbf{(IV)}}}
\end{overpic}
\caption{Fits to the $M(\gamma X)$ invariant mass distributions, where
  $X$ represents (I) $\pi^{+}\pi^{-}$, (II) $\pi^{+}\pi^{-}\eta$,
  (III) $2(\pi^{+}\pi^{-})$, and (IV) $p\bar{p}$. The dots
  with error bars are data, the red solid line is the fitting result,
  the pink dashed line is the signal shape, the blue dashed line is
  the smooth background and the other dashed lines represent the peaking
  backgrounds. }
\label{fig:fit_hc}
\end{center}
\end{figure}

The signal yields are determined by fits to the $M(\gamma X)$
invariant mass distributions, which are shown in Fig.~\ref{fig:fit_hc}
for all decay modes. The signal shape is
described by the MC-simulated shape convolved with a Gaussian
function with free parameters to account for the resolution
difference between data and MC. The peaking backgrounds are fixed to
the estimated numbers and described by the corresponding MC
shapes. The smooth background is modeled by an ARGUS
function~\cite{argus}, where the threshold parameter is fixed to 3.551
GeV/$c^2$. The statistical significance of each decay mode is larger
than 5$\,\sigma$, which is estimated by comparing the log-likelihood
values of the fits with and without the signal component and considering
the change of degrees of freedom.

\begin{figure}[htbp]
\begin{center}
\begin{overpic}[width=0.42\textwidth, trim=50 30 20 20]{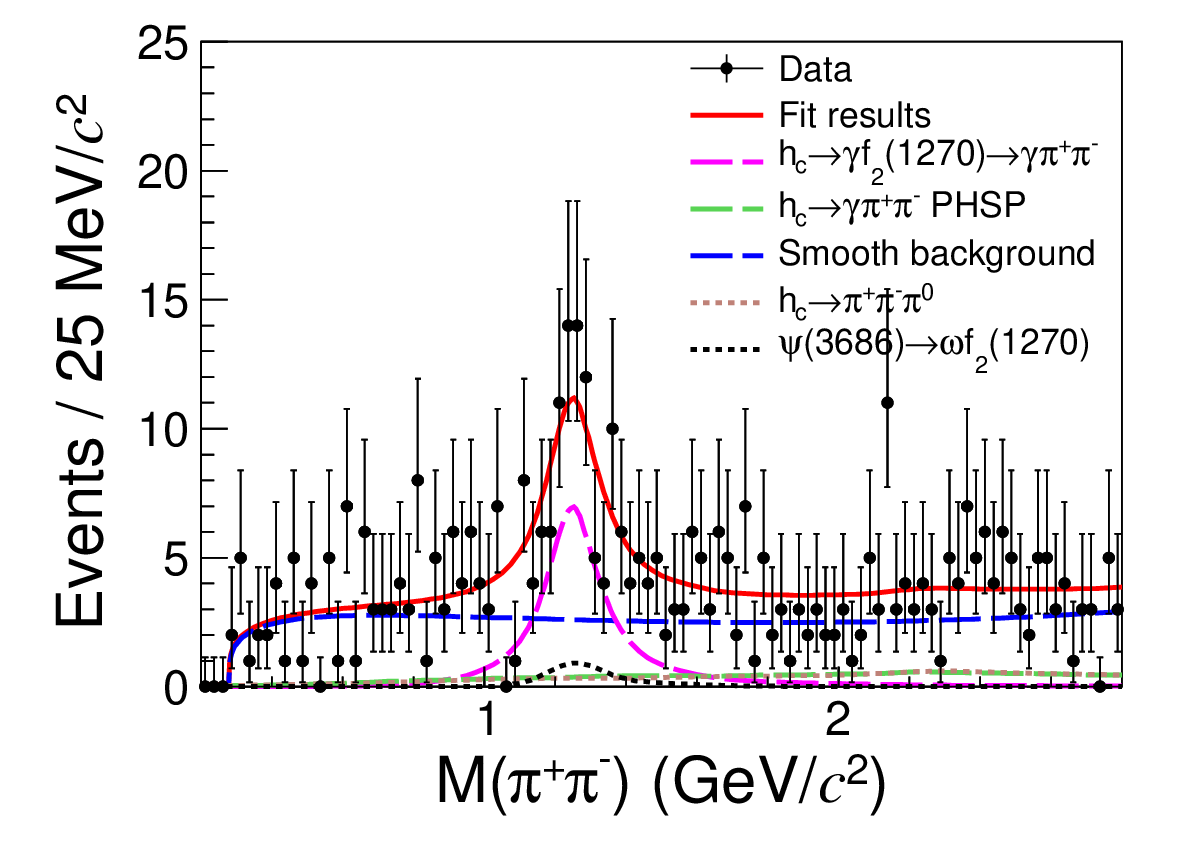}
\end{overpic}
\begin{overpic}[width=0.42\textwidth, trim=50 30 20 20]{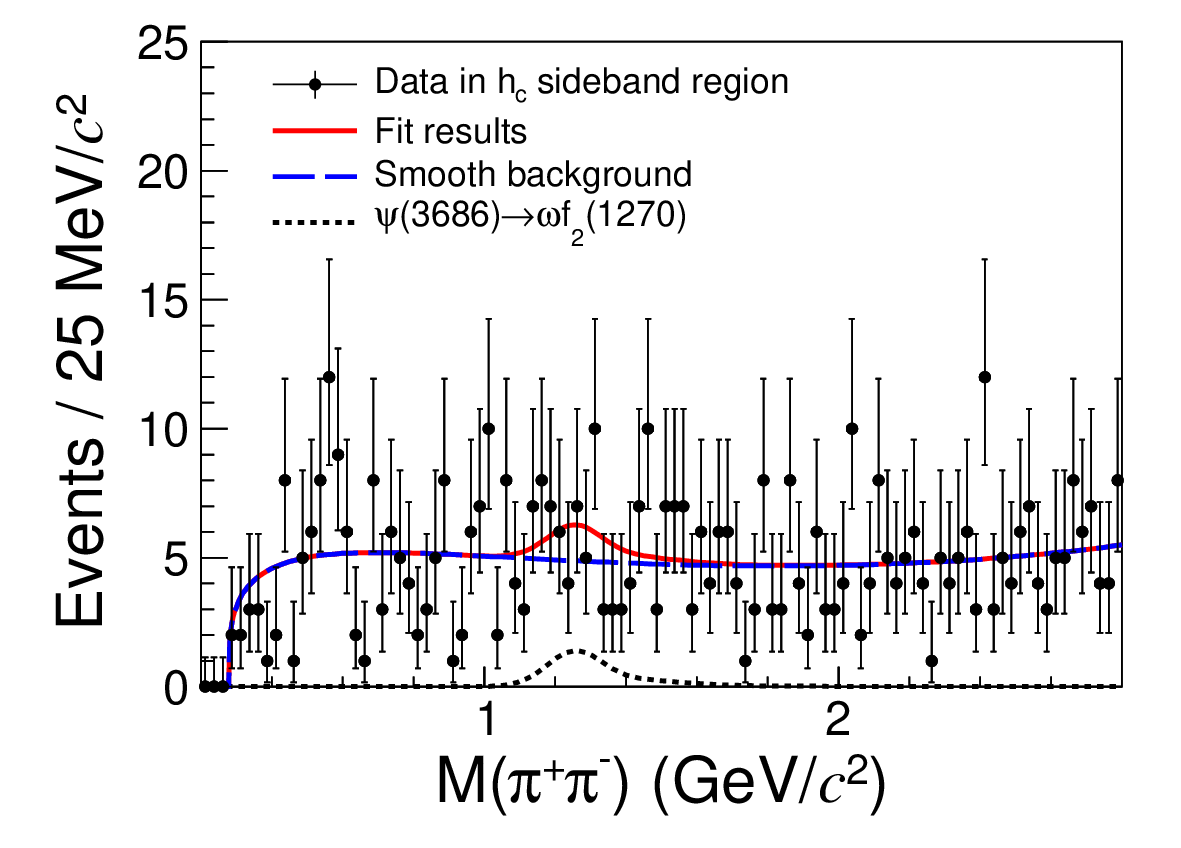}
\end{overpic}
\caption{Fit to the $M(\pi^{+}\pi^{-})$ invariant mass 
  distribution for the process of $h_{c}\too\gamma\pi^{+}\pi^{-}$ for
  events in the (top) $h_c$ signal region and (bottom) $h_c$ sideband
  regions.  The dots with error bars are data, the red solid line is
  the total fitting result, the pink dashed line is the signal of
  $h_{c}\too\gamma f_{2}(1270)$ and the green dashed line is the phase
  space process of $h_{c}\too\gamma\pi^{+}\pi^{-}$. The blue dashed
  line denotes the smooth background, and the brown dashed line the
  background from $h_{c}\too\pi^{+}\pi^{-}\pi^{0}$. The black dashed
  line is the $\psi(3686) \to \omega f_2(1270)$ peaking background.  }
\label{fig:f2_2pi}
\end{center}
\end{figure}

To search for possible intermediate states, the invariant mass
distributions of light hadrons are examined within the $h_c$ signal
region [3.52, 3.53] GeV/$c^{2}$. Figure~\ref{fig:f2_2pi} (top) shows
the $M(\pi^{+}\pi^{-})$ distribution of
$h_{c}\too\gamma\pi^{+}\pi^{-}$ below 2.8 GeV/$c^{2}$ from this region. An excess of
events around 1.3 GeV/$c^{2}$ is observed, suggesting a potential
resonance, assumed to be the $f_{2}(1270)$ state. The statistical
significance of the $f_2(1270)$ signal is determined to be
5.5$\,\sigma$, and after including systematic uncertainties (see below), the
significance is 5.1$\,\sigma$.

To determine the number of $f_{2}(1270)$ signal events, we perform a
simultaneous fit to the $\pi^+\pi^-$ mass spectra in the $h_c$ signal
region and sideband regions ([3.50, 3.51]$\cup$[3.54, 3.55]
GeV/$c^{2}$).  Included in the fit of the signal region are the $h_c
\to \gamma f_2(1270)$ signal process, non-resonant $h_c \to \gamma \pi^+\pi^-$ described by
phase space, and the following background processes: $\psi(3686) \to
\omega f_2(1270)$ peaking background, $h_c \to \pi^+\pi^-\pi^0$, and a
smooth background.  The shapes of these components are described by MC
shapes, except the $f_{2}(1270)$ signal is described by its
MC shape convolved with a Gaussian
function with free parameters, and the smooth background is described
by a function $F_{BG}(m)=(m-m_{t})^{c}e^{-d\cdot m-e\cdot m^{2}}$,
where $m_{t}$ is the threshold mass for $\pi^{+}\pi^{-}$ and $c,~d$,
and $e$ are free parameters. The number of $\psi(3686) \to
\omega f_2(1270)$ peaking
background events is fixed to 9.0 according to a MC study, and the
number of background events from $h_c\too\pi^{+}\pi^{-}\pi^{0}$ is
fixed to 35.4 according to Ref.~\cite{hc3pi}. 
Two potential peaking background processes, $\psi(3686) \to \gamma \pi^0 f_2(1270)$ and $\psi(3686) \to \pi^0 f_2(1270)$, have been investigated. The former is evaluated using $h_c$ sideband data and found to be negligible. The latter, being forbidden by C-parity conservation, is expected to be highly suppressed and thus neglected in this analysis.

The fit to the sideband includes the $\psi(3686) \to \omega f_2(1270)$
peaking background, described by the MC shape, and the smooth
background, described by $F_{BG}(m)$.  The two fit
functions share the smooth background $F_{BG}(m)$, and the number of
smooth background events in the $h_{c}$ signal region is scaled by a
scale factor from the number of smooth background events in the
$h_{c}$ sideband region.  The number of phase space and signal events
are floated in the fitting process, but the interference between them
is not considered due to the low statistics.  The fits in the $h_c$
signal and sideband regions are shown in Fig.~\ref{fig:f2_2pi}.

The branching fractions of $h_c\too\gamma X$ ($X=\pi^{+}\pi^{-}$,
$\pi^{+}\pi^{-}\eta$, $2(\pi^{+}\pi^{-})$, $p\bar{p}$) are calculated
by
\begin{equation*}
\begin{aligned}
    \mathcal{B}(h_c\too\gamma X) =\frac{N^{\rm sig}}{N^{\rm{tot}}\cdot{\mathcal{B}(\psi(3686)\too \pi^{0}h_c)\prod_{i}\mathcal{B}_{i}\epsilon}},
    \end{aligned}
\end{equation*}
where $N^{\rm sig}$ is the number of signal events,
$\prod_{i}\mathcal{B}_{i}$ is the product of branching fractions of
the intermediate states,
e.g. $\mathcal{B}(\piz\too\gamma\gamma)$~\cite{pdg} or
$\mathcal{B}(\piz\too\gamma\gamma)\cdot\mathcal{B}(\eta\too\gamma\gamma)$~\cite{pdg},
$\epsilon$ is the detection efficiency, and $N^{\rm{tot}}$ is the
number of $\psi(3686)$ events. For the $h_c\too\gamma\pi^{+}\pi^{-}$
process, the total detection efficiency is estimated by combining the
PHSP and intermediate state $f_{2}(1270)$ processes. The ratio between
these two components is determined based on this study.

 For $h_{c}\too\gamma\pi^{+}\pi^{-}\eta$ and $h_{c}\too\gamma p
 \bar{p}$, no obvious intermediate state is found. We find that
 $h_c\too\gamma2(\pi^{+}\pi^{-})$ is dominated by
 $h_{c}\too\gamma\rho^{0}\rho^{0}$, therefore,
 $h_c\too\gamma2(\pi^{+}\pi^{-})$ is simulated by
 $h_{c}\too\gamma\rho^{0}\rho^{0}\too\gamma2(\pi^{+}\pi^{-})$ cascade
 decay. Table~\ref{tab:brsum} summarizes the branching fraction,
 signal yield, efficiency, and significance for each decay mode.

\begin{table*}[htbp]
  \centering
   \small
  \caption{The numerical results for four $h_c$ decay channels, where
    the first uncertainty is statistical, the second systematic, and
    the third from the branching fraction of
    $\psi(3686)\too\pi^{0}h_c$. The significance is estimated
    taking systematic uncertainties into account.}
  \label{tab:brsum}
  \begin{minipage}[t]{0.95\textwidth}
  \centering
  \begin{tabular}{lcccc}
  \hline
  \hline
  Decay model   \ \ \ \  &   \ \ \ \  Branching fraction  \ \ \ \  & \ \ \ \  Signal yield \ \  \ \  &  \ \ \ \ Efficiency (\%)  \ \ \ \  &  \ \ \ \ Significance ($\sigma$)   \\
  \hline
  $h_c\too\gamma\pi^{+}\pi^{-}$   & $(3.06\pm0.54\pm0.37\pm0.21)\times10^{-4}$ & $127.0\pm22.2$ & 20.9 & 5.4 \\
  $h_c\too\gamma f_{2}(1270)\too\gamma\pi^{+}\pi^{-}$ & $(1.81\pm0.35\pm0.19\pm0.12)\times10^{-4}$ & $71.8\pm13.8$ & 20.0 & 5.1 \\
  $h_c\too\gamma\pi^{+}\pi^{-}\eta$   &  $(3.52\pm0.50\pm0.30\pm0.24)\times10^{-3}$ & $191.6\pm27.1$ & 7.0 & 7.5 \\
  $h_c\too\gamma2(\pi^{+}\pi^{-})$  & $(2.19\pm0.20\pm0.16\pm0.15)\times10^{-3}$ & $494.9\pm44.5$ &11.4  & 11.8\\
  $h_c\too\gamma p\bar{p}$  &  $(3.34\pm0.53\pm0.33\pm0.23)\times10^{-4}$ & $127.4\pm20.1$ & 19.3 & 6.4 \\
  \hline
  \hline
  \end{tabular}
\end{minipage}
\end{table*}

Systematic uncertainties of these decay branching fractions are listed
in Table~\ref{tab:summererror}. The overall systematic uncertainties are
obtained by adding all sources in quadrature assuming
they are independent.

\begin{table*}[htbp]
\centering
\caption{Relative systematic uncertainties (\%), where a dash(-) indicates that a systematic effect is not applicable.}
\label{tab:summererror}
\begin{tabular}{lcccccccc}
\hline
\hline
Source  \ \ \  & $h_c\too\gamma\pi^{+}\pi^{-}$ &  \ \ \ \   $h_c\too\gamma\pi^{+}\pi^{-}\eta$ \ \ \ \  & \ \ \ \  $h_c\too\gamma2(\pi^{+}\pi^{-})$ &  \ \ \ \  $h_c\too\gamma p\bar{p}$ \ \ \ \  &  \ \  \  \ $h_{c}\too\gamma f_{2}(1270)\too\gamma\pi^{+}\pi^{-}$ \\
\hline
Tracking                     & 2.0  & 2.0    &  4.0 &  1.7 &   2.0  \\
Photon reconstruction        & 1.5  & 2.5    & 1.5  &  1.5 &  1.5  \\
PID                          & -    &  -     & -    &  2.9 &  -  \\
$\pi^{0}$  reconstruction    & 0.5  & 0.5    & 0.5  &  0.5  & 0.5 \\
$\eta$ mass window           & -    & 1.0    & -    &  -    & -  \\
Kinematic fit                & 1.5  & 0.6    & 2.1  & 0.6  & 1.3 \\
Intermediate states          &2.0   & 0.1    &0.6   &-     &-\\
Background veto              &2.0   & 2.2    &-     &-     &2.0\\
Peaking background           &8.7   & 4.3    &1.5   & 0.6   & 2.7 \\
Fitting procedure            &7.5   &6.0     &5.3   & 9.1   &9.7\\
Branching fractions           &6.8   &6.8     &6.8   & 6.8   & 6.8 \\
Number of $\psi(3686)$ & 0.5  & 0.5    & 0.5  & 0.5   & 0.5 \\
\hline
Total                        & 14.0 & 10.9   & 10.0 & 12.0  & 12.7 \\
\hline
\hline
\end{tabular}
\end{table*}

The uncertainty arising from the pion tracking efficiency is $1.0\%$
per track according to the study of a control sample of
$\psi(3686)\too\pi^{+}\pi^{-}J/\psi$, $J/\psi\too
\ell^{+}\ell^{-}~(\ell=e,~\mu)$~\cite{track}. Similarly, the tracking
efficiencies of protons and anti-protons are studied with the control
sample of $\psi(3686)\too p\bar{p}\pi^{+}\pi^{-}$. The uncertainty
associated with the proton(anti-proton) tracking efficiency is 0.7\%
(1.0\%) per track~\cite{trackproton}. The uncertainty due to PID is
1.3\% per proton and 1.6\% per anti-proton, based on the study of
$J/\psi\too p\bar{p}\pi^{+}\pi^{-}$~\cite{hc3pi}.

The uncertainty of the photon detection efficiency is $0.5\%$ per
photon obtained from the study of the $\EE\too\gamma\mu^{+}\mu^{-}$
process. The uncertainty of the $\pi^{0}$ reconstruction efficiency is
estimated using its momentum distribution based on the control sample
of $\psi(3686)\too\piz\piz J/\psi$ and $\EE\too\omega\piz$ at
$\sqrt{s}$=3.773 GeV. MC events are weighted according to the relative
difference between data and MC, which is described by a linear function
($0.06-2.41\times p$)\%, where $p$ is the momentum of $\pi^0$. The
uncertainty due to the $\eta$ mass window is determined to be 1\%
through a study of the control sample of $J/\psi\too
p\bar{p}\eta$~\cite{track}.

The uncertainty due to the kinematic fit requirements is estimated by
correcting the helix parameters of charged tracks according to the
method described in Ref.~\cite{kinematic}. The difference between
detection efficiencies obtained from MC samples with and without this
correction is taken as the uncertainty.

Possible intermediate states are considered in the simulations of
$h_c\too\gamma\pi^{+}\pi^{-}$, $h_c\too\gamma\pi^{+}\pi^{-}\eta$, and
$h_c\too\gamma2(\pi^{+}\pi^{-})$ modes. For
$h_c\too\gamma\pi^{+}\pi^{-}$, the detection efficiency is estimated
based on the ratio of the PHSP and intermediate state
$f_{2}(1270)$ processes. The covariance matrix obtained from this
study is used to regenerate multi-dimensional Gaussian samples to get
the uncertainty of the average efficiency. For
$h_c\too\gamma\pi^{+}\pi^{-}\eta$, we reweighted the PHSP MC sample
according to distributions of $M(\pi^{+}\pi^{-})$, $M(\pi^{+}\eta)$,
and $M(\pi^{-}\eta)$ in the data. In the case of
$h_c\too\gamma2(\pi^{+}\pi^{-})$, the distribution of
$M(\pi^{+}\pi^{-})$ in the data is used to weight the $h_{c}\too\gamma\rho^0\rho^0$ 3-body PHSP MC. The efficiency
differences between the PHSP MC and the weighted MC are taken as
uncertainties.

For the $\pi^{0}$ background veto, the photon energy resolution difference
between data and MC can affect the distribution of
$M(\gamma\gamma)$. To address this, we adjust the energy resolution by
4\%~\cite{gamres}, and the difference between the efficiencies with or
without this correction is regarded as the systematic uncertainty.  For
the uncertainty caused by other background vetoes, we perform a Barlow
test~\cite{barlow} to examine the significance deviation ($\zeta$)
between the nominal fit and the systematic test. We change the
background veto region in 10 different intervals. If $\zeta$ is less
than 2 at all times (or very close to 2), we treat the systematic
uncertainty as negligible. Otherwise, we take the maximum difference
as the systematic uncertainty for a conservative estimation.

There are three sources of peaking background. For the $\psi(3686) \to
\omega f_2(1270)$
and $h_c\rightarrow\pi^{0}X$ peaking backgrounds, the systematic
uncertainties are estimated based on the uncertainties of measured
branching fractions~\cite{pdg, hc3pi, hc3pieta}. For
$h_c\rightarrow\gamma\eta_{c}$, the fitting results are varied within
one standard deviation, and the differences between the varied and nominal
values are taken as the systematic uncertainties.

The uncertainty from the fitting procedure is determined by varying
the fitting range, background shape, and signal shape. The uncertainty
caused by the fitting range is estimated by changing the left side of
the range by $\pm10~{\rm MeV}/c^{2}$~\cite{hcgamP,hc3pi}. The difference in the fitting
results is taken as the systematic uncertainty. To estimate the
systematic uncertainty related to the background shape, the ARGUS
function is changed to a function $(m_{t}-m)^{a}e^{-bm-cm^{2}}$~\cite{truncated}, where $m_{t}=3.551\,{\rm GeV}/c^{2}$, $a$, $b$ and $c$ are free parameters. The uncertainty of the signal shape is estimated
by replacing the nominal fit by a Breit-Wigner function convolved with
a Gaussian function with free parameters, where the parameters of
Breit-Wigner function are fixed to the values in the
PDG~\cite{pdg}. The uncertainty in the fitting procedure is determined
by the differences in the final results added in quadrature.

For the $h_c\too\gamma f_{2}(1270)\too\gamma\pi^{+}\pi^{-}$ process, the
uncertainty of the fitting procedure involves the background from
$h_{c}\too\pi^{+}\pi^{-}\pi^{0}$, the $h_{c}$ signal region and $h_{c}$
sideband regions, and the $h_{c}$ sideband scale factor. The uncertainty
of the background from $h_{c}\too\pi^{+}\pi^{-}\pi^{0}$ is
estimated by varying the background contribution within its
uncertainty. To estimate the uncertainties related to the choice of
signal and sideband regions, the $h_{c}$ signal region is varied by
$\pm1~{\rm MeV}/c^{2}$ at both boundaries and the $h_{c}$ sideband
regions are shifted to the left or right by $\pm2~{\rm MeV}/c^{2}$.
The $h_{c}$ sideband scale factor is determined according to the fit
results of $M(\gamma\pi^{+}\pi^{-})$. To estimate the uncertainty, we
recalculate the scale factor using the covariance matrix from the fit
results. Five thousand sets of fitting parameters are generated using
multi-dimensional Gaussian sampling to recalculate the $h_{c}$
sideband scaling factor. The standard deviation of the resultant
number of $f_{2}(1270)$ events is taken as the systematic uncertainty.
The uncertainty in the fitting procedure is determined
by differences in the final results added in quadrature.
  
The systematic uncertainties arising from the branching fractions,
including $\mathcal{B}(\psi(3686)\too \pi^{0}h_c)$,
$\mathcal{B}(\eta\too\gamma\gamma)$, and
$\mathcal{B}(\piz\too\gamma\gamma)$, are from the PDG~\cite{pdg}.
The number of $\psi(3686)$ is determined by measuring the inclusive
hadronic events, as described in Ref.~\cite{0912data}. The uncertainty
is estimated to be 0.5\%.

In summary, using a data sample of $(27.12\pm0.14)\times10^{8}$
$\psi(3686)$ events collected by the BESIII detector, four modes of
$h_c$ radiative decays to multiple light hadrons are studied via
$\psi(3686)\too\pi^{0}h_{c}$ ($h_c\too\gamma X$) for the first
time. The processes of $h_c\too\gamma X~(X=\pi^{+}\pi^{-}$,
$\pi^{+}\pi^{-}\eta$, $2(\pi^{+}\pi^{-})$, and $p\bar{p})$ are
observed with a significance larger than 5$\,\sigma$. 
The sum of known branching fractions of the $h_c$ radiative decay including the results of this analysis is approximately 0.8\%, which is significantly lower than the
predicted value 5.5\% of $\mathcal{B}(h_{c}\too\gamma +2g)$ in
Ref.~\cite{qcdhc}. This discrepancy highlights the need for further
exploration of $h_{c}$ radiative decays. Moreover, $h_c$ radiative
decay to $f_{2}(1270)$ is observed with a significance of
$5.1\,\sigma$ including the systematic uncertainty. This is the first
observation of $h_c$ radiative decay to a tensor state. The branching
fraction of $h_c\too\gamma f_{2}(1270)\too\gamma\pi^{+}\pi^{-}$ is an
order of magnitude smaller than that in $J/\psi$
decay~\cite{jpsigamf2}. This measurement could be employed to
investigate the quark and glueball mixing parameters of $f_{2}(1270)$
from the method as outlined in Ref.~\cite{qqgmodel}.

\begin{acknowledgments}
The BESIII Collaboration thanks the staff of BEPCII and the IHEP computing center for their strong support. This work is supported in part by National Key R\&D Program of China under Contracts Nos. 2020YFA0406300, 2020YFA0406400, 2023YFA1606000; National Natural Science Foundation of China (NSFC) under Contracts Nos. 12375070, 11635010, 11735014, 11935015, 11935016, 11935018, 12025502, 12035009, 12035013, 12061131003, 12192260, 12192261, 12192262, 12192263, 12192264, 12192265, 12221005, 12225509, 12235017, 12361141819; the Chinese Academy of Sciences (CAS) Large-Scale Scientific Facility Program; the CAS Center for Excellence in Particle Physics (CCEPP); Shanghai Leading Talent Program of Eastern Talent Plan under Contract No. JLH5913002; Joint Large-Scale Scientific Facility Funds of the NSFC and CAS under Contract No. U2032108, U1832207; 100 Talents Program of CAS; The Institute of Nuclear and Particle Physics (INPAC) and Shanghai Key Laboratory for Particle Physics and Cosmology; German Research Foundation DFG under Contract No. FOR5327; Istituto Nazionale di Fisica Nucleare, Italy; Knut and Alice Wallenberg Foundation under Contracts Nos. 2021.0174, 2021.0299; Ministry of Development of Turkey under Contract No. DPT2006K-120470; National Research Foundation of Korea under Contract No. NRF-2022R1A2C1092335; National Science and Technology fund of Mongolia; National Science Research and Innovation Fund (NSRF) via the Program Management Unit for Human Resources \& Institutional Development, Research and Innovation of Thailand under Contracts Nos. B16F640076, B50G670107; Polish National Science Centre under Contract No. 2019/35/O/ST2/02907; Swedish Research Council under Contract No. 2019.04595; The Swedish Foundation for International Cooperation in Research and Higher Education under Contract No. CH2018-7756; U. S. Department of Energy under Contract No. DE-FG02-05ER41374
\end{acknowledgments}

\end{document}